\documentclass[]{aa}  

\usepackage{graphicx}
\usepackage{txfonts}

\begin{document}

   \title{Detection of glycolaldehyde towards the solar-type protostar NGC~1333~IRAS2A\thanks{Based on observations carried out with the IRAM Plateau de Bure Interferometer. IRAM is supported by INSU/CNRS (France), MPG (Germany) and IGN (Spain).}}

   \author{A. Coutens\inst{1} \and M. V. Persson\inst{2} \and J. K. J\o rgensen\inst{1} \and S. F. Wampfler\inst{1} \and J. M. Lykke\inst{1}
          }

   \institute{Centre for Star and Planet Formation, Niels Bohr Institute and Natural History Museum of Denmark, University of Copenhagen, \O ster Voldgade 5-7, DK-1350 Copenhagen K, Denmark\\
              \email{acoutens@nbi.dk}
                    \and
 	Leiden Observatory, Leiden University, PO Box 9513, 2300 RA Leiden, The Netherlands            
             }

   \date{Received xxx; accepted xxx}
 
  \abstract
  {Glycolaldehyde is a key molecule in the formation of biologically relevant molecules such as ribose. We report its detection with the Plateau de Bure interferometer towards the Class 0 young stellar object NGC~1333~IRAS2A, which is only the second solar-type protostar for which this prebiotic molecule is detected. Local thermodynamic equilibrium analyses of glycolaldehyde, ethylene glycol (the reduced alcohol of glycolaldehyde) and methyl formate (the most abundant isomer of glycolaldehyde) were carried out. The relative abundance of ethylene glycol to glycolaldehyde is found to be $\sim$5 -- higher than in the Class 0 source IRAS~16293-2422 ($\sim$1), but comparable to the lower limits derived in comets ($\ge$~3--6). 
The different ethylene glycol-to-glycolaldehyde ratios in the two protostars could be related to different CH$_3$OH:CO compositions of the icy grain mantles. In particular, a more efficient hydrogenation on the grains in NGC~1333~IRAS2A would favor the formation of both methanol and ethylene glycol.
In conclusion, it is possible that, like NGC~1333~IRAS2A, other low-mass protostars show high ethylene glycol-to-glycolaldehyde abundance ratios. The cometary ratios could consequently be inherited from earlier stages of star formation, if the young Sun experienced conditions similar to NGC~1333~IRAS2A.
}


   \maketitle
%

\section{Introduction}

The inner regions of low-mass protostars are known to harbor a rich complex organic chemistry characterized by the presence of molecules such as methyl formate (CH$_3$OCHO), dimethyl ether (CH$_3$OCH$_3$), and ethyl cyanide (C$_2$H$_5$CN) \citep[e.g.,][]{Cazaux2003,Bottinelli2004a,Bisschop2008}. To differentiate them from the hot cores present in high-mass star-forming regions, they were called hot corinos \citep{Ceccarelli2004,Bottinelli2004b}. These complex organic molecules are thought to be efficiently formed on grains and then released into the gas phase in the hot corino by thermal desorption \citep[e.g.,][]{Garrod2008,Herbst2009}. 
Some of these complex organic molecules are particularly interesting because of their supposed role in the emergence of life. Indeed, the detection of so called prebiotic molecules in low-mass star-forming regions indicates that they can form early during the star formation process and thereby be available for possible later incorporation into solar system bodies, e.g., comets. 

Glycolaldehyde (CH$_2$OHCHO) is one of these prebiotic molecules: it is a simple sugar-like molecule and under Earth-like conditions the first product in the formose reaction leading to the formation of ribose, an essential constituent of ribonucleic acid (RNA) \citep[e.g.,][]{Zubay2001,Jalbout2007}. Glycolaldehyde was first detected towards the Galactic center (Sgr B2(N): \citealt{Hollis2000,Hollis2001,Hollis2004,Halfen2006,Belloche2013}; molecular clouds: \citealt{Requena2008}). Later it was shown to be present in the high mass star-forming region G31.41+0.31 \citep{Beltran2009}, in the intermediate mass protostar NGC~7129~FIRS~2 \citep{Fuente2014}, and even in the hot corinos of the Class 0 protostellar binary, IRAS~16293-2422 \citep[hereafter IRAS16293,][]{Jorgensen2012}. This indicates that this molecule can be synthesized relatively early in the environments of solar-type protostars. 
Furthermore, glycolaldehyde can easily survive during impact delivery to planetary bodies, and impacts can even facilitate the formation of even more complex molecules \citep{McCaffrey2014}.

Similarly to other complex organic molecules, the formation of glycolaldehyde is thought to occur on grains. In particular, a gas phase formation was excluded by \citet{Woods2012,Woods2013}, as the produced abundances are too low compared with the observations. Several grain surface formation pathways were proposed in the literature. \citet{Woods2012} modeled their efficiency and showed that the formation by the reaction CH$_3$OH + HCO would be very efficient, but that, from chemical considerations, H$_3$CO + HCO could be more feasible. Another probably efficient way to form glycolaldehyde would be through HCO dimerization (HCO + HCO $\rightarrow$ HOCCOH) followed by two successive hydrogenations \citep{Woods2013}. A recent experimental study based on surface hydrogenations of CO  seems to confirm this pathway \citep{Fedoseev2014}.

A related species to this prebiotic molecule is ethylene glycol ((CH$_2$OH)$_2$). More commonly known as antifreeze, it is the reduced alcohol of glycolaldehyde. 
This molecule was tentatively detected towards IRAS16293 with one line of the gGg$'$ conformer \citep{Jorgensen2012}. Interestingly, the aGg$'$ conformer of ethylene glycol (the conformer of lowest energy) was detected in three comets, Hale-Bopp, Lemmon, and Lovejoy, while glycolaldehyde was not, leading to a lower limit of 3--6 for the (CH$_2$OH)$_2$/CH$_2$OHCHO abundance ratio \citep{Crovisier2004,Biver2014}. Ethylene glycol was also detected in the Murchison and Murray carbonaceous meteorites, while the presence of aldehyde sugars have not been reported yet \citep{Cooper2001}. 

NGC~1333~IRAS2A (hereafter IRAS2A) is another of these famous hot corinos. In particular, methyl formate, the most abundant isomer of glycolaldehyde, was detected towards this source by \citet{Jorgensen2005a} and \citet{Bottinelli2007}. More recently, ethylene glycol was detected in the framework of the CALYPSO program carried out with the IRAM Plateau de Bure Interferometer (PdBI) by \citet{Maury2014}. We here report the detection of glycolaldehyde towards the same low-mass protostar, and present
 an analysis of the relative abundances of these three species.


\section{Observations}
\label{sect_obs}

This work is based on several separate programs carrying out observations of the solar-type protostar IRAS2A with the PdBI. Four spectral ranges (84.9$-$88.5, 223.5$-$227.1, 240.2$-$243.8, and 315.5$-$319.1 GHz) were covered with the WIDEX correlator at a spectral resolution of 1.95 MHz ($d\varv$ = 6.8 km\,s$^{-1}$ at 86 GHz, $d\varv$ = 2.6 km\,s$^{-1}$ at 225 GHz, $d\varv$ = 2.4 km\,s$^{-1}$ at 242 GHz, $d\varv$ = 1.8 km\,s$^{-1}$ at 317 GHz) and reduced with the GILDAS\footnote{\url{http://www.iram.fr/IRAMFR/GILDAS/}} software. The synthetized beam sizes obtained with natural weighting are about 3.0$\arcsec$\,$\times$\,3.0$\arcsec$ at 86 GHz, 1.2$\arcsec$\,$\times$\,1.0$\arcsec$ at 225 GHz, 1.4$\arcsec$\,$\times$\,1.0$\arcsec$ at 242 GHz, and 0.9$\arcsec$\,$\times$\,0.8$\arcsec$ at 317 GHz. The dust continuum fluxes at 0.9 and 1.3 mm are consistent with previous measurements \citep[e.g.,][]{Jorgensen2007,Persson2012}. The absolute calibration uncertainty for each dataset is about 20\%. 
Additional information about the observations and their reduction can be found in \citet{Coutens2014} and \citet{Persson2014}. The 3mm data are from Wampfler (priv. comm.).

Using the CASSIS\footnote{\url{http://cassis.irap.omp.eu}} software, we detected 8 lines of glycolaldehyde, 31 lines of the aGg$'$ conformer of ethylene glycol, and 26 lines of methyl formate (see Table \ref{table_obs}). The glycolaldehyde and methyl formate transitions are taken from the JPL spectroscopic database \citep{Pickett1998}, while the ethylene glycol transitions are from the CDMS catalogue \citep{Muller2001,Muller2005}. The predictions are based on experimental data from \citet{Butler2001}, \citet{Widicus2005} and \citet{Carroll2010} for glycolaldehyde, \citet{Christen1995} and \citet{Christen2003} for ethylene glycol, and \citet{Ilyushin2009} for methyl formate.
The frequencies of five of the detected glycolaldehyde lines were directly measured in the laboratory \citep{Butler2001}.
Some of the lines result from a blending of several transitions of the same species. The lines that are strongly blended with other species are not listed in Table\,\ref{table_obs}.  
All three species are emitted very compactly at the position of the continuum peak ($\alpha_{2000}$=$03^{\rm h}28^{\rm m}55\fs57$, $\delta_{2000}$=$31\degr14\arcmin37\farcs1$). The angular sizes obtained with a circular Gaussian fit in the ($u$,$\varv$) plane vary from a point source to a maximum of 1$\arcsec$ depending on the transition. The line fluxes listed in Table \ref{table_obs} were measured at the continuum peak position with the CASSIS software using a Gaussian fitting method (Levenberg-Marquardt algorithm). The lines that are contaminated in the wings by other transitions are consequently fitted with a sum of Gaussians. We carefully checked that the derived full widths at half maximum ($FWHM$) are consistent with the other line measurements. The average  $FWHM$ is about 4.5 km\,s$^{-1}$ at 317 GHz, and 5.0 km\,s$^{-1}$ at 225 and 242 GHz. The widths of the methyl formate lines at 87 GHz are quite broad ($\sim$12 km\,s$^{-1}$). It is consequently difficult to completely exclude an extra flux contribution from other species. The variation of FWHM with the frequency can be explained by the spectral resolution of the observations that decreases towards the lower frequencies.

\begin{longtab}
\vspace{-0.2cm}
\begin{longtable}{llrrcrclcc}
\caption{CH$_2$OHCHO, aGg$'$-(CH$_2$OH)$_2$ and CH$_3$OCHO transitions observed towards NGC~1333~IRAS2A. }
\label{table_obs}
 \\
\hline\hline	
Species & Transition & Frequency  & $E_{\rm up}$ & $A_{\rm ij}$ & $g_{\rm up}$ & Flux  & RD$^a$ \\
& & (MHz) & (K) &(s$^{-1}$) &  & (Jy km s$^{-1}$) \\
\hline
CH$_2$OHCHO & 13$_{10,3}-$13$_{9,4}$ ($\varv$=0) &  240366.34$^*$ & 111.3 & 1.2\,$\times$\,10$^{-4}$ & 27 & 0.055 & Y \\
 & 13$_{10,4}-$13$_{9,5}$ ($\varv$=0) &  240366.34$^*$ & 111.3 & 1.2\,$\times$\,10$^{-4}$ & 27 & \\
CH$_2$OHCHO & 12$_{10,2}-$12$_{9,3}$ ($\varv$=0) &  240482.78$^*$ & 104.0 & 1.0\,$\times$\,10$^{-4}$ & 25 & 0.038  & Y \\ 
 & 12$_{10,3}-$12$_{9,4}$ ($\varv$=0) &  240482.78$^*$ & 104.0 & 1.0\,$\times$\,10$^{-4}$ & 25 & \\ 
CH$_2$OHCHO & 11$_{5,6}-$10$_{4,7}$ ($\varv$=0) &  240890.46$\,$~ & 51.9 & 1.8\,$\times$\,10$^{-4}$ & 23 & 0.663 & N$^b$ \\ 
CH$_2$OHCHO & 22$_{2,20}-$21$_{3,19}$ ($\varv$=0) &  241131.84$\,$~ & 142.8 & 2.8\,$\times$\,10$^{-4}$ & 45 & 0.061 & Y \\ 
CH$_2$OHCHO & 23$_{2,22}-$22$_{1,21}$ ($\varv$=0) & 242239.09$\,$~ & 146.2 & 3.5\,$\times$\,10$^{-4}$ & 47 & 0.128 & Y \\
CH$_2$OHCHO & 24$_{0,24}-$23$_{1,23}$ ($\varv$=0) & 242957.72$^*$ & 148.2 & 4.2\,$\times$\,10$^{-4}$ & 49 & 0.258 & Y \\
 & 24$_{1,24}-$23$_{0,23}$ ($\varv$=0) &  242957.98$^*$ & 148.2 & 4.2\,$\times$\,10$^{-4}$ & 49 & \\
CH$_2$OHCHO & 19$_{13,7}-$19$_{12,8}$ ($\varv$=0) & 315941.48$^*$ & 208.1 & 3.2\,$\times$\,10$^{-4}$ & 39 & 0.074 & Y \\
 & 19$_{13,6}-$19$_{12,7}$ ($\varv$=0) & 315941.48$^*$ & 208.1 & 3.2\,$\times$\,10$^{-4}$ & 39 & \\
CH$_2$OHCHO & 11$_{8,4}-$10$_{7,3}$ ($\varv$=0) & 317013.88$^*$ & 75.5 & 6.6\,$\times$\,10$^{-4}$ & 23 & 0.363 & Y$^c$ \\
 & 11$_{8,3}-$10$_{7,4}$ ($\varv$=0) & 317013.90$^*$ & 75.5 & 6.6\,$\times$\,10$^{-4}$ & 23 & \\
CH$_2$OHCHO & 27$_{5,23}-$26$_{4,22}$ ($\varv$=0) & 317850.44$\,$~ & 226.2 & 4.5\,$\times$\,10$^{-4}$ & 55 & 0.117 & Y$^c$ \\
\hline
aGg$'$-(CH$_2$OH)$_2$ & 21$_{6,16}$ ($\varv$=1)--20$_{6,15}$ ($\varv$=0) & 223741.66$\,$~ & 132.0 & 2.5\,$\times$\,10$^{-4}$ & 387 & 0.212 & Y \\
aGg$'$-(CH$_2$OH)$_2$ & 21$_{6,15}$ ($\varv$=1)--20$_{6,14}$ ($\varv$=0) & 224405.85$\,$~ & 132.1 & 2.5\,$\times$\,10$^{-4}$ & 301 & 0.197 & Y \\
aGg$'$-(CH$_2$OH)$_2$ & 24$_{0,24}$ ($\varv$=1)--23$_{1,23}$ ($\varv$=0) & 224511.70$^*$ & 136.8 & 5.4\,$\times$\,10$^{-5}$ & 441 & 0.145 & N \\
& 24$_{1,24}$ ($\varv$=1)--23$_{0,23}$ ($\varv$=0) & 224512.74$^*$ & 136.8 & 5.4\,$\times$\,10$^{-5}$ & 343 & \\
aGg$'$-(CH$_2$OH)$_2$ & 21$_{3,18}$ ($\varv$=1)--20$_{3,17}$ ($\varv$=0) & 225688.94$\,$~ & 121.3 & 2.4\,$\times$\,10$^{-4}$ & 301 & 0.195 & Y \\
aGg$'$-(CH$_2$OH)$_2$ & 22$_{3,20}$ ($\varv$=1)--21$_{3,19}$ ($\varv$=0) & 225929.69$\,$~ & 127.8 & 2.5\,$\times$\,10$^{-4}$ & 315 & 0.349 & N$^c$ \\
aGg$'$-(CH$_2$OH)$_2$ & 22$_{5,17}$ ($\varv$=0)--21$_{5,16}$ ($\varv$=1) & 226095.96$\,$~ & 138.2 & 2.6\,$\times$\,10$^{-4}$ & 315 & 0.303 & N \\
aGg$'$-(CH$_2$OH)$_2$ & 22$_{2,20}$ ($\varv$=1)--21$_{2,19}$ ($\varv$=0) & 226561.99$\,$~ & 127.7 & 3.0\,$\times$\,10$^{-4}$ & 405 & 0.329 & Y \\
aGg$'$-(CH$_2$OH)$_2$ & 25$_{1,25}$ ($\varv$=0)--24$_{1,24}$ ($\varv$=1) & 226643.30$^*$ & 147.7 & 2.9\,$\times$\,10$^{-4}$ & 357 & 0.470 & Y \\
& 25$_{0,25}$ ($\varv$=0)--24$_{0,24}$ ($\varv$=1) & 226643.46$^*$ & 147.7 & 2.9\,$\times$\,10$^{-4}$ & 459 &  \\
aGg$'$-(CH$_2$OH)$_2$  & 25$_{1,25}$ ($\varv$=1)--24$_{1,24}$ ($\varv$=0) & 240778.12$^*$ & 148.0 & 3.4\,$\times$\,10$^{-4}$ & 459 & 0.360 & Y \\
& 25$_{0,25}$ ($\varv$=1)--24$_{0,24}$ ($\varv$=0) & 240778.30$^*$ & 148.0 & 3.4\,$\times$\,10$^{-4}$ & 357 & \\
aGg$'$-(CH$_2$OH)$_2$  & 24$_{8,17}$ ($\varv$=0)--23$_{8,16}$ ($\varv$=1) & 240807.88$\,$~ & 179.2 & 3.0\,$\times$\,10$^{-4}$ & 441 & 0.340 & N$^e$ \\
aGg$'$-(CH$_2$OH)$_2$  & 24$_{8,16}$ ($\varv$=0)--23$_{8,15}$ ($\varv$=1) & 240828.89$\,$~ & 179.2 & 3.0\,$\times$\,10$^{-4}$ & 343 & 0.149 & Y \\
aGg$'$-(CH$_2$OH)$_2$  & 24$_{5,20}$ ($\varv$=0)--23$_{5,19}$ ($\varv$=1) & 241291.27$\,$~ & 160.7 & 3.1\,$\times$\,10$^{-4}$ & 441 & 0.196 & Y \\
aGg$'$-(CH$_2$OH)$_2$  & 24$_{7,18}$ ($\varv$=0)--23$_{7,17}$ ($\varv$=1) & 241545.26$\,$~ & 172.1 & 3.1\,$\times$\,10$^{-4}$ & 441 & 0.151 & Y \\
aGg$'$-(CH$_2$OH)$_2$  & 24$_{6,19}$ ($\varv$=0)--23$_{6,18}$ ($\varv$=1) & 241860.73$\,$~ & 166.0 & 2.8\,$\times$\,10$^{-4}$ & 441 & 0.137 & Y \\
aGg$'$-(CH$_2$OH)$_2$ & 23$_{15,8}$ ($\varv$=1)--22$_{15,7}$ ($\varv$=0) & 242244.69$^*$ & 246.4 & 2.0\,$\times$\,10$^{-4}$ & 329 & 0.474 & N \\
& 23$_{15,9}$ ($\varv$=1)--22$_{15,8}$ ($\varv$=0) & 242244.69$^*$ & 246.4 & 2.0\,$\times$\,10$^{-4}$ & 423 & \\
& 23$_{6,17}$ ($\varv$=1)--22$_{6,17}$ ($\varv$=1) & 242245.62$^*$ & 154.6 & 1.1\,$\times$\,10$^{-5}$ & 329 & \\
& 23$_{14,9}$ ($\varv$=1)--22$_{14,8}$ ($\varv$=0) & 242246.34$^*$ & 232.2 & 2.2\,$\times$\,10$^{-4}$ & 329 & \\
& 23$_{14,10}$ ($\varv$=1)--22$_{14,9}$ ($\varv$=0) & 242246.34$^*$ & 232.2 & 2.2\,$\times$\,10$^{-4}$ & 423 & \\
aGg$'$-(CH$_2$OH)$_2$ & 23$_{13,10}$ ($\varv$=1)--22$_{13,9}$ ($\varv$=0) & 242277.72$^*$  & 218.9 & 2.4\,$\times$\,10$^{-4}$ & 329 & 0.275 & N \\
& 23$_{13,11}$ ($\varv$=1)--22$_{13,10}$ ($\varv$=0) & 242277.72$^*$  & 218.9 & 2.4\,$\times$\,10$^{-4}$ & 423 & \\
aGg$'$-(CH$_2$OH)$_2$ & 23$_{10,14}$ ($\varv$=1)--22$_{10,13}$ ($\varv$=0) & 242656.22$^*$ & 185.2 & 2.8\,$\times$\,10$^{-4}$ & 423 & 0.305 & Y \\
& 23$_{10,13}$ ($\varv$=1)--22$_{10,12}$ ($\varv$=0) & 242656.24$^*$  & 185.2 & 2.8\,$\times$\,10$^{-4}$ & 329 &  \\
aGg$'$-(CH$_2$OH)$_2$ & 23$_{9,15}$ ($\varv$=1)--22$_{9,14}$ ($\varv$=0) & 242947.99$^*$  & 175.9 & 3.0\,$\times$\,10$^{-4}$ & 423 & 0.291 & Y \\
& 23$_{9,14}$ ($\varv$=1)--22$_{9,13}$ ($\varv$=0) & 242948.59$^*$  & 175.9 & 3.0\,$\times$\,10$^{-4}$ & 329 &  \\
aGg$'$-(CH$_2$OH)$_2$ & 23$_{5,19}$ ($\varv$=1)--22$_{5,18}$ ($\varv$=0) & 243636.57$^*$  & 149.1 & 3.4\,$\times$\,10$^{-4}$ & 423 & 0.287 & Y \\
aGg$'$-(CH$_2$OH)$_2$ & 31$_{8,23}$ ($\varv$=0)--30$_{8,22}$ ($\varv$=1) & 315671.33$\,$~ & 276.6 & 7.0\,$\times$\,10$^{-4}$ & 567 & 0.314 & N$^c$ \\
aGg$'$-(CH$_2$OH)$_2$ & 31$_{7,25}$ ($\varv$=0)--30$_{7,24}$ ($\varv$=1) & 315892.11$\,$~ & 269.6 & 6.8\,$\times$\,10$^{-4}$ & 441 & 0.144 & N$^d$ \\
aGg$'$-(CH$_2$OH)$_2$ & 30$_{9,21}$ ($\varv$=1)--29$_{9,20}$ ($\varv$=0) & 315961.89$\,$~ & 269.4 & 7.0\,$\times$\,10$^{-4}$ & 549 & 0.259 & Y \\
aGg$'$-(CH$_2$OH)$_2$ & 34$_{1,34}$ ($\varv$=0)--33$_{0,33}$ ($\varv$=0) & 316444.07$^*$ & 268.5 & 1.6\,$\times$\,10$^{-4}$ & 621 & 0.078 & Y \\
 & 34$_{0,34}$ ($\varv$=0)--33$_{1,33}$ ($\varv$=0) & 316444.07$^*$ & 268.5 & 1.6\,$\times$\,10$^{-4}$ & 483 & \\
aGg$'$-(CH$_2$OH)$_2$ & 20$_{6,14}$ ($\varv$=0)--19$_{5,15}$ ($\varv$=0) & 316698.08$\,$~ & 121.3 & 5.1\,$\times$\,10$^{-5}$ & 287 & 0.029 & Y \\
aGg$'$-(CH$_2$OH)$_2$ & 30$_{7,24}$ ($\varv$=1)--29$_{7,23}$ ($\varv$=0) & 316868.23$^*$ & 254.4 & 6.9\,$\times$\,10$^{-4}$ & 427 & 0.445 & N$^c$ \\ 
 & 20$_{16,4}$ ($\varv$=0)--20$_{15,5}$ ($\varv$=0) & 316870.92$^*$ & 228.9 & 3.8\,$\times$\,10$^{-5}$ & 287 \\
 & 20$_{16,5}$ ($\varv$=0)--20$_{15,6}$ ($\varv$=0) & 316870.92$^*$ & 228.9 & 3.8\,$\times$\,10$^{-5}$ & 369 \\
aGg$'$-(CH$_2$OH)$_2$ & 30$_{8,23}$ ($\varv$=1)--29$_{8,22}$ ($\varv$=0) & 316917.19$\,$~ & 261.4 & 7.2\,$\times$\,10$^{-4}$ & 427 & 0.299 & Y \\
aGg$'$-(CH$_2$OH)$_2$ & 16$_{8,9}$ ($\varv$=1)--15$_{7,8}$ ($\varv$=1) & 317054.30$^*$ & 98.6 & 8.0\,$\times$\,10$^{-5}$ & 231 & 0.116 & N \\
 & 16$_{8,8}$ ($\varv$=1)--15$_{7,9}$ ($\varv$=1) & 317055.36$^*$ & 98.6 & 8.0\,$\times$\,10$^{-5}$ & 297 \\
aGg$'$-(CH$_2$OH)$_2$ & 21$_{4,18}$ ($\varv$=1)--20$_{4,18}$($\varv$=0) & 317267.77$^*$ & 122.1 & 1.1\,$\times$\,10$^{-5}$ & 387 & 0.312 & N \\
& 14$_{9,5}$ ($\varv$=1)--13$_{8,6}$ ($\varv$=1) & 317267.91$^*$ & 91.4 & 1.1\,$\times$\,10$^{-4}$ & 261 \\
 & 14$_{9,6}$ ($\varv$=1)--13$_{8,5}$ ($\varv$=1) & 317267.91$^*$ & 91.4 & 1.1\,$\times$\,10$^{-4}$ & 203 \\
 & 14$_{9,5}$ ($\varv$=0)--13$_{8,6}$ ($\varv$=0) & 317268.88$^*$ & 91.4 & 1.1\,$\times$\,10$^{-4}$ & 203 \\
 & 14$_{9,6}$ ($\varv$=0)--13$_{8,5}$ ($\varv$=0) & 317268.88$^*$ & 91.4 & 1.1\,$\times$\,10$^{-4}$ & 261 \\
aGg$'$-(CH$_2$OH)$_2$ & 32$_{3,30}$ ($\varv$=1)--31$_{3,29}$ ($\varv$=0) & 317962.58$\,$~ & 257.2 & 7.8\,$\times$\,10$^{-4}$ & 455 & 0.310 & Y \\
aGg$'$-(CH$_2$OH)$_2$ & 32$_{4,28}$ ($\varv$=0)--31$_{4,27}$ ($\varv$=1) & 317982.56$\,$~ & 272.2 & 4.8\,$\times$\,10$^{-4}$ & 455 & 0.160 & Y \\
aGg$'$-(CH$_2$OH)$_2$ & 35$_{0,35}$ ($\varv$=0)--34$_{0,34}$ ($\varv$=1) & 318433.40$^*$ & 284.1 & 8.0\,$\times$\,10$^{-4}$ & 639 & 0.639 & N \\ 
 & 35$_{1,35}$ ($\varv$=0)--34$_{1,34}$ ($\varv$=1) & 318433.40$^*$ & 284.1 & 8.0\,$\times$\,10$^{-4}$ & 497 \\ 
\hline 
CH$_3$OCHO & 7$_{3,4}$--6$_{3,3}$ E ($\varv_T$=0) & 87143.28$^*$ & 22.6 & 7.7\,$\times$\,10$^{-6}$ & 30 & 0.049 & N\\
& 21$_{5,16}$--21$_{4,17}$ E ($\varv_T$=1) & 87143.65$^*$ & 342.0 & 1.4\,$\times$\,10$^{-6}$ & 86 \\
CH$_3$OCHO & 8$_{0,8}$--7$_{1,7}$ E ($\varv_T$=1) & 87160.84$^*$ & 207.0 & 1.3\,$\times$\,10$^{-6}$ & 34 & 0.038 & N \\
& 7$_{3,4}$--6$_{3,3}$ A ($\varv_T$=0) & 87161.28$^*$ & 22.6 & 7.8\,$\times$\,10$^{-6}$ & 30 \\
CH$_3$OCHO & 18$_{6,13}$--17$_{6,12}$ E ($\varv_T$=0) & 224021.87$^*$ & 125.3 & 1.5\,$\times$\,10$^{-4}$ & 74 & 0.752 & Y \\
& 18$_{6,13}$--17$_{6,12}$ A ($\varv_T$=0) & 224024.10$^*$ & 125.3 & 1.5\,$\times$\,10$^{-4}$ & 74 \\
CH$_3$OCHO & 18$_{5,14}$--17$_{5,13}$ E ($\varv_T$=0) & 224313.15$\,$~ & 118.3 & 1.6\,$\times$\,10$^{-4}$ & 74 & 0.351 & Y \\
CH$_3$OCHO & 19$_{3,17}$--18$_{3,16}$ E ($\varv_T$=1) & 224491.31$\,$~ & 303.2 & 1.7\,$\times$\,10$^{-4}$ & 78 & 0.212 & Y \\
CH$_3$OCHO & 18$_{6,12}$--18$_{6,11}$ E ($\varv_T$=0) & 224582.35$\,$~ & 125.4 & 1.5\,$\times$\,10$^{-4}$ & 74 & 0.463 & Y \\
CH$_3$OCHO & 18$_{6,12}$--17$_{6,11}$ A ($\varv_T$=0) & 224609.38$\,$~  & 125.4 & 1.5\,$\times$\,10$^{-4}$ & 74 & 0.466 & Y \\ 
CH$_3$OCHO & 20$_{2,19}$--19$_{2,18}$ A ($\varv_T$=1) & 225372.22$\,$~ & 307.3 & 1.7\,$\times$\,10$^{-4}$ & 82 & 0.186 & Y \\
CH$_3$OCHO & 19$_{3,17}$--18$_{3,16}$ E ($\varv_T$=0) & 225608.82$\,$~ & 116.7 & 1.7\,$\times$\,10$^{-4}$ & 78 & 0.426 & Y \\
CH$_3$OCHO & 18$_{5,13}$--17$_{5,12}$ A ($\varv_T$=1) & 225648.42$^*$ & 305.6 & 1.6\,$\times$\,10$^{-4}$ & 74 & 0.163 & N \\
& 26$_{9,18}$--26$_{8,19}$ A ($\varv_T$=0) & 225648.42$^*$ & 261.7 & 1.6\,$\times$\,10$^{-5}$ & 106   \\
CH$_3$OCHO & 21$_{0,21}$--20$_{1,20}$ A ($\varv_T$=1) & 226381.36$^*$ & 309.6 & 2.8\,$\times$\,10$^{-5}$ & 86 & 0.277 & Y \\
& 21$_{1,21}$--20$_{1,20}$ A ($\varv_T$=1) & 226382.72$^*$ & 309.6 & 1.7\,$\times$\,10$^{-4}$ & 86 \\
& 21$_{0,21}$--20$_{0,20}$ A ($\varv_T$=1) & 226383.86$^*$ & 309.6 & 1.7\,$\times$\,10$^{-4}$ & 86 \\
& 21$_{1,21}$--20$_{0,20}$ A ($\varv_T$=1) & 226385.15$^*$ & 309.6 & 2.8\,$\times$\,10$^{-5}$ & 86 \\
CH$_3$OCHO & 21$_{0,21}$--20$_{1,20}$ E ($\varv_T$=1) & 226433.26$^*$ & 308.9 & 2.7\,$\times$\,10$^{-5}$ & 86 & 0.296 & N \\
& 21$_{1,21}$--20$_{1,20}$ E ($\varv_T$=1) & 226434.47$^*$ & 308.9 & 1.7\,$\times$\,10$^{-4}$ & 86 \\
& 21$_{0,21}$--20$_{0,20}$ E ($\varv_T$=1) & 226435.52$^*$ & 308.9 & 1.7\,$\times$\,10$^{-4}$ & 86 \\
& 25$_{9,16}$--25$_{8,17}$ A ($\varv_T$=0) & 226435.52$^*$ & 246.2 & 1.6\,$\times$\,10$^{-5}$ & 102 \\
& 21$_{1,21}$--20$_{0,20}$ E ($\varv_T$=1) & 226436.66$^*$ & 308.9 & 2.7\,$\times$\,10$^{-5}$ & 86 \\
CH$_3$OCHO & 20$_{2,19}$--19$_{2,18}$ A ($\varv_T$=0)  & 226718.69$\,$~ & 120.2 & 1.7\,$\times$\,10$^{-4}$ & 82 & 0.505 & Y  \\
CH$_3$OCHO & 19$_{2,17}$--18$_{2,16}$ E ($\varv_T$=0) & 227019.55$^*$ & 116.6 & 1.7\,$\times$\,10$^{-4}$ & 78 & 0.561 & N \\
 & 25$_{9,17}$--25$_{8,18}$ A ($\varv_T$=0) & 227021.13$^*$ & 246.2 & 1.6\,$\times$\,10$^{-5}$ & 102 \\
CH$_3$OCHO & 19$_{2,17}$--18$_{2,16}$ A ($\varv_T$=0) & 227028.12$\,$~ & 116.6 & 1.7\,$\times$\,10$^{-4}$ & 78 & 0.436 & Y \\
CH$_3$OCHO & 20$_{4,17}$--19$_{4,16}$ A ($\varv_T$=1) & 242610.07$\,$~ & 321.7 & 2.1\,$\times$\,10$^{-4}$ & 82 & 0.092 & Y \\
CH$_3$OCHO & 37$_{7,31}$--37$_{6,32}$ A ($\varv_T$=0) & 242870.39$^*$ & 452.0 & 1.8\,$\times$\,10$^{-5}$ & 150 & 0.714 & N\\
 & 19$_{5,14}$--18$_{5,13}$ E ($\varv_T$=0) & 242871.57$^*$ & 130.5 & 2.0\,$\times$\,10$^{-4}$ & 78 \\
CH$_3$OCHO  & 19$_{5,14}$--18$_{5,13}$ A ($\varv_T$=0) & 242896.02$\,$~ & 130.4 & 2.0\,$\times$\,10$^{-4}$ & 78 & 0.666 & Y \\
CH$_3$OCHO & 21$_{12,9}$--21$_{11,10}$ E ($\varv_T$=0) & 316742.00$^*$  & 231.8 & 3.3\,$\times$\,10$^{-5}$ & 86 & 0.072 & Y \\
 & 21$_{12,10}$--21$_{11,11}$ E ($\varv_T$=0) & 316742.71$^*$  & 231.8 & 3.3\,$\times$\,10$^{-5}$ & 86 & \\
CH$_3$OCHO &  21$_{12,9}$--21$_{11,10}$ A ($\varv_T$=0) & 316776.74$^*$ & 231.8 & 3.3\,$\times$\,10$^{-5}$ & 86 & 0.069 & Y \\
 &  21$_{12,10}$--21$_{11,11}$ A ($\varv_T$=0) & 316776.74$^*$ & 231.8 & 3.3\,$\times$\,10$^{-5}$ & 86 \\
CH$_3$OCHO & 26$_{13,13}$--25$_{13,12}$ E ($\varv_T$=1) & 317177.16$\,$~ & 506.5 & 3.6\,$\times$\,10$^{-4}$ & 106 & 0.043 & Y \\
CH$_3$OCHO & 9$_{8,2}$--8$_{7,1}$ A ($\varv_T$=1) & 318009.06$^{*}$ & 256.8 & 6.8\,$\times$\,10$^{-5}$ & 38 & 0.200 & N \\
 & 9$_{8,1}$--8$_{7,2}$ A ($\varv_T$=1) & 318009.06$^{*}$ & 256.8 & 6.9\,$\times$\,10$^{-5}$ & 38 \\
 & 13$_{12,1}$--13$_{11,2}$ E ($\varv_T$=0) & 318009.55$^{*}$ & 149.2 & 1.3\,$\times$\,10$^{-5}$ & 54 \\
 & 26$_{13,13}$--25$_{13,12}$ A ($\varv_T$=1) & 318012.17$^{*}$ & 506.2 & 3.7\,$\times$\,10$^{-4}$ & 106 \\
 & 26$_{13,14}$--25$_{13,13}$ A ($\varv_T$=1) & 318012.17$^{*}$ & 506.2 & 3.7\,$\times$\,10$^{-4}$ & 106 \\
CH$_3$OCHO  & 13$_{12,2}$--13$_{11,3}$ E ($\varv_T$=0) & 318016.90$^{*}$ & 149.2 &  1.3\,$\times$\,10$^{-5}$ & 54 & 0.155 & N \\
 & 9$_{8,1}$--8$_{7,1}$ E ($\varv_T$=0) & 318017.37$^{*}$ & 69.0 & 6.7\,$\times$\,10$^{-5}$ & 38 \\
CH$_3$OCHO & 12$_{12,0}$--12$_{11,1}$ E ($\varv_T$=0) & 318064.54$^{*}$ & 141.6 & 7.0\,$\times$\,10$^{-6}$ & 50 & 0.204 & N\\
 & 9$_{8,2}$--8$_{7,1}$ A ($\varv_T$=0) & 318065.26$^{*}$  & 69.0 & 6.8\,$\times$\,10$^{-5}$ & 38 \\
 & 9$_{8,1}$--8$_{7,2}$ A ($\varv_T$=0) & 318065.26$^{*}$  & 69.0 & 6.8\,$\times$\,10$^{-5}$ & 38 \\
CH$_3$OCHO & 27$_{4,23}$--26$_{5,22}$ E ($\varv_T$=1) & 318139.11$^{*}$ & 426.4 & 3.4\,$\times$\,10$^{-5}$ & 110 & 0.192 & N\\
 & 25$_{6,19}$--24$_{6,18}$ A ($\varv_T$=1) & 318140.72$^{*}$  & 405.1 & 4.6\,$\times$\,10$^{-4}$ & 102 \\
 & 26$_{11,15}$--25$_{11,14}$ E ($\varv_T$=1) & 318145.25$^{*}$ & 474.5 & 4.0\,$\times$\,10$^{-4}$ & 106 \\
CH$_3$OCHO & 27$_{4,24}$--26$_{4,23}$ E ($\varv_T$=1) & 318979.14$\,$~ & 417.8 & 4.8\,$\times$\,10$^{-4}$ & 110 & 0.208 & Y \\
\hline
\end{longtable}
{ Notes: The symbol $^*$ present after some frequency values indicates that the associated transition is blended with one or more transitions from the same species. $^a$Y indicates that the line was considered in the rotational diagram analysis, while N indicates that it could not be used (for blending reasons). $^b$Blended with an unidentified species. $^c$Blended with CH$_3$OCHO. $^d$Blended with CH$_2$OHCHO. $^e$Potentially blended with the gGg$'$ conformer of ethylene glycol.}
\end{longtab}


\section{Results}
\label{sect_analysis}

\begin{figure}[t!]
\begin{center}
\includegraphics[scale=0.65]{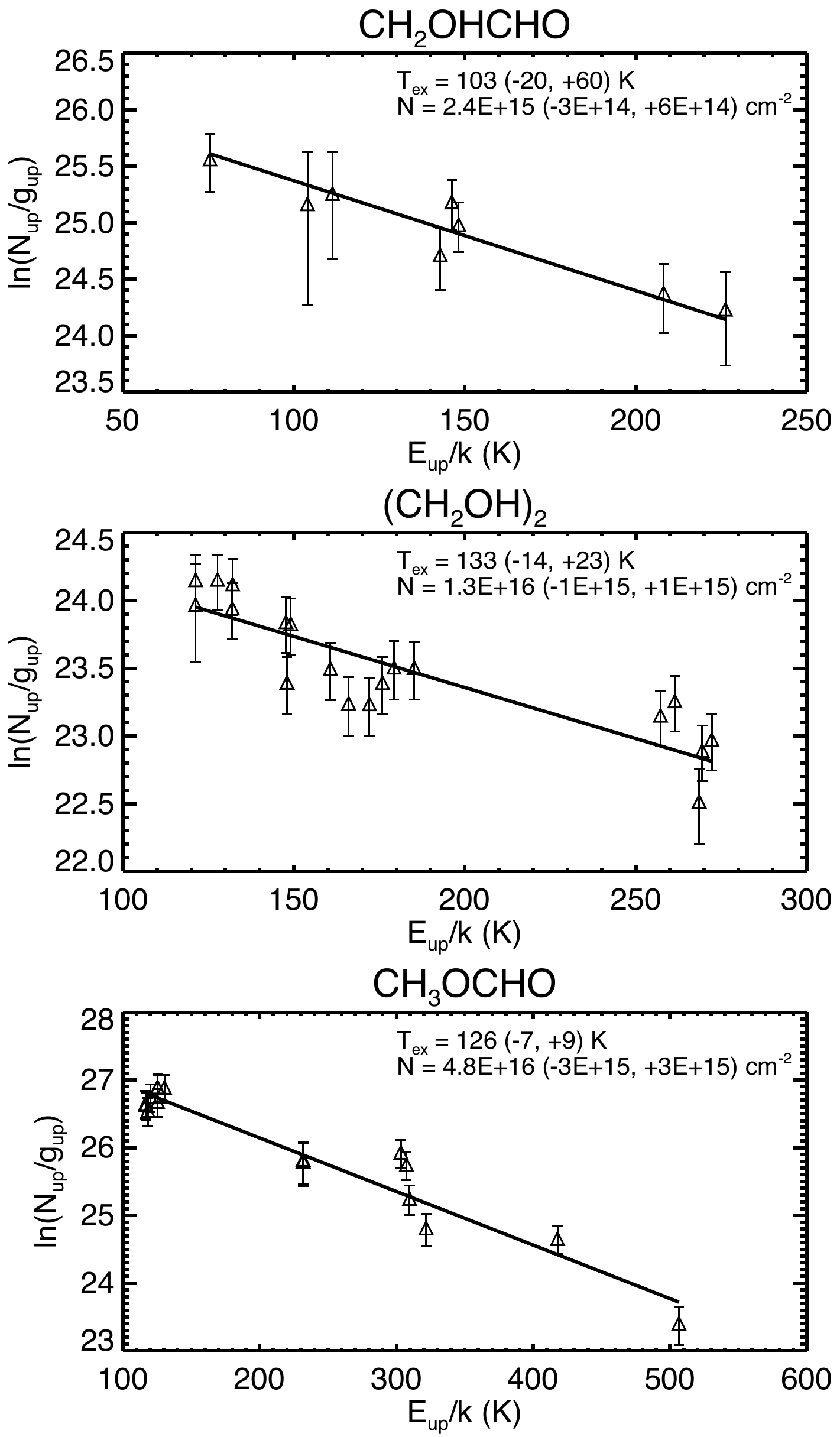}
\caption{Rotational diagrams for glycolaldehyde, ethylene glycol, and methyl formate.} 
\label{Fig_RD}
\end{center}
\end{figure}

\begin{figure*}[t!]
\begin{center}
\includegraphics[scale=0.95]{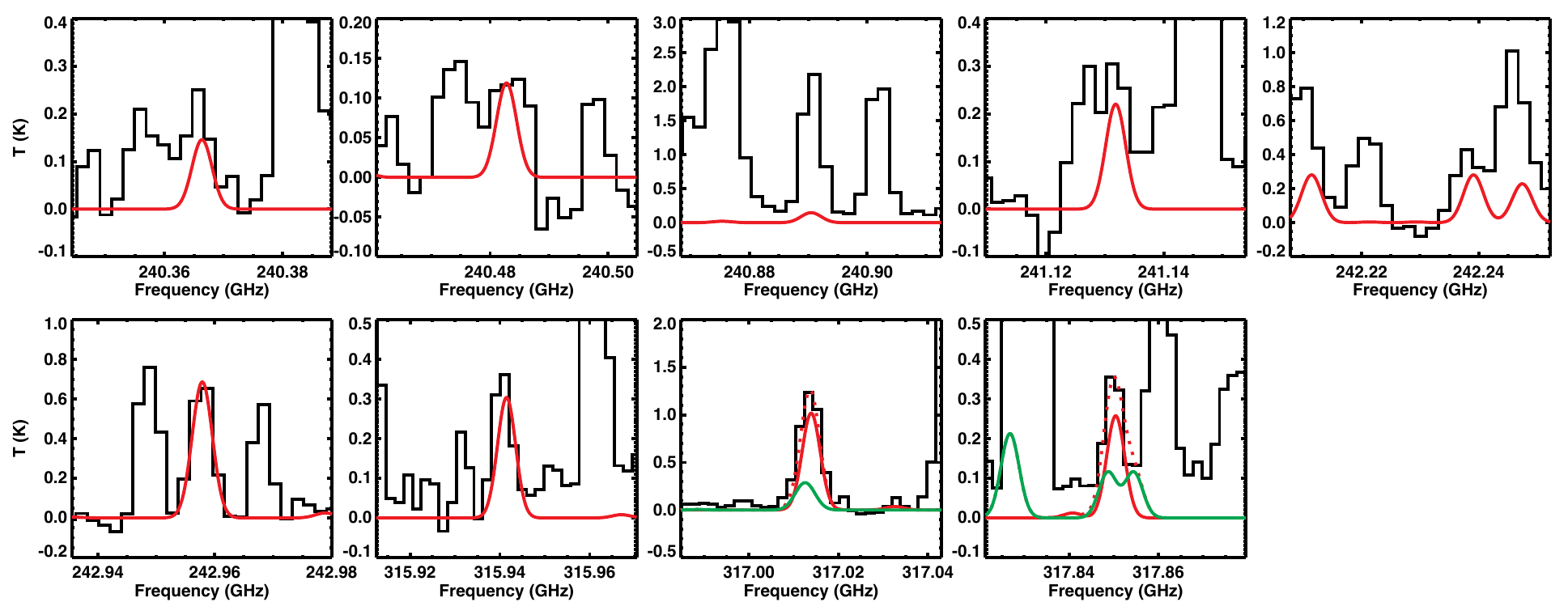}
\caption{Observed lines of CH$_2$OHCHO towards the protostar NGC~1333~IRAS2A (in black). The best-fit model for CH$_2$OHCHO (see Table \ref{table_model}) is shown in red solid lines. The line in the third upper panel is blended with an unidentified species. The contribution of the CH$_3$OCHO lines is indicated with green lines, and the model including both CH$_2$OHCHO and CH$_3$OCHO can be seen in red dotted lines.}
\label{Model_GA}
\end{center}
\end{figure*}
\begin{figure*}[h!]
\begin{center}
\includegraphics[scale=0.95]{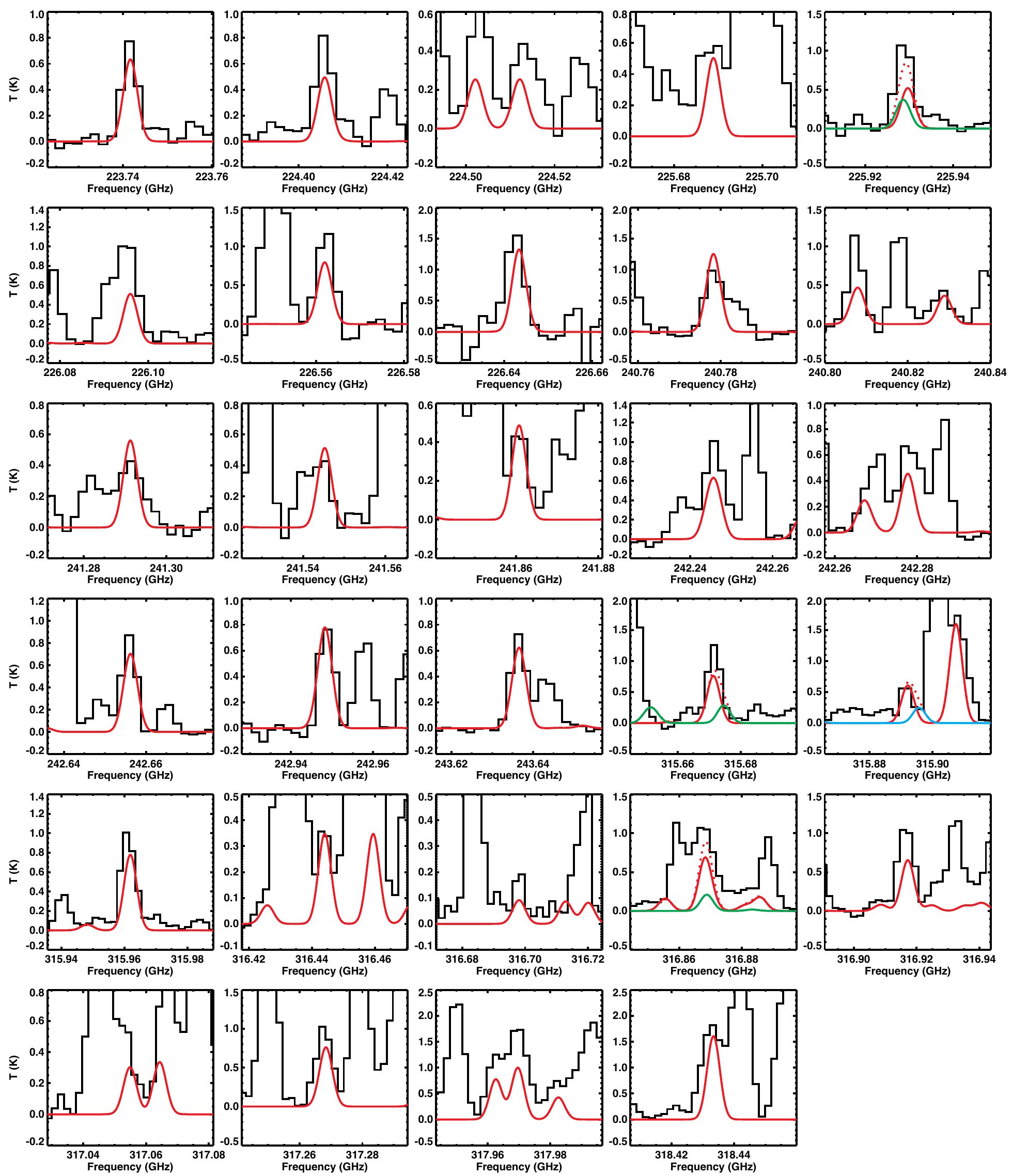}
\caption{
Observed lines of aGg$'$-(CH$_2$OH)$_2$ towards the protostar NGC~1333~IRAS2A (in black). The best-fit model for aGg$'$-(CH$_2$OH)$_2$ (see Table \ref{table_model}) is shown in red solid lines. The contributions of the CH$_3$OCHO and CH$_2$OHCHO lines are indicated with green and blue lines, respectively. The model including aGg$'$-(CH$_2$OH)$_2$, CH$_2$OHCHO, and CH$_3$OCHO can be seen in red dotted lines.}
\label{Model_aGgglycol}
\end{center}
\end{figure*}
\begin{figure*}[h!]
\begin{center}
\includegraphics[scale=0.95]{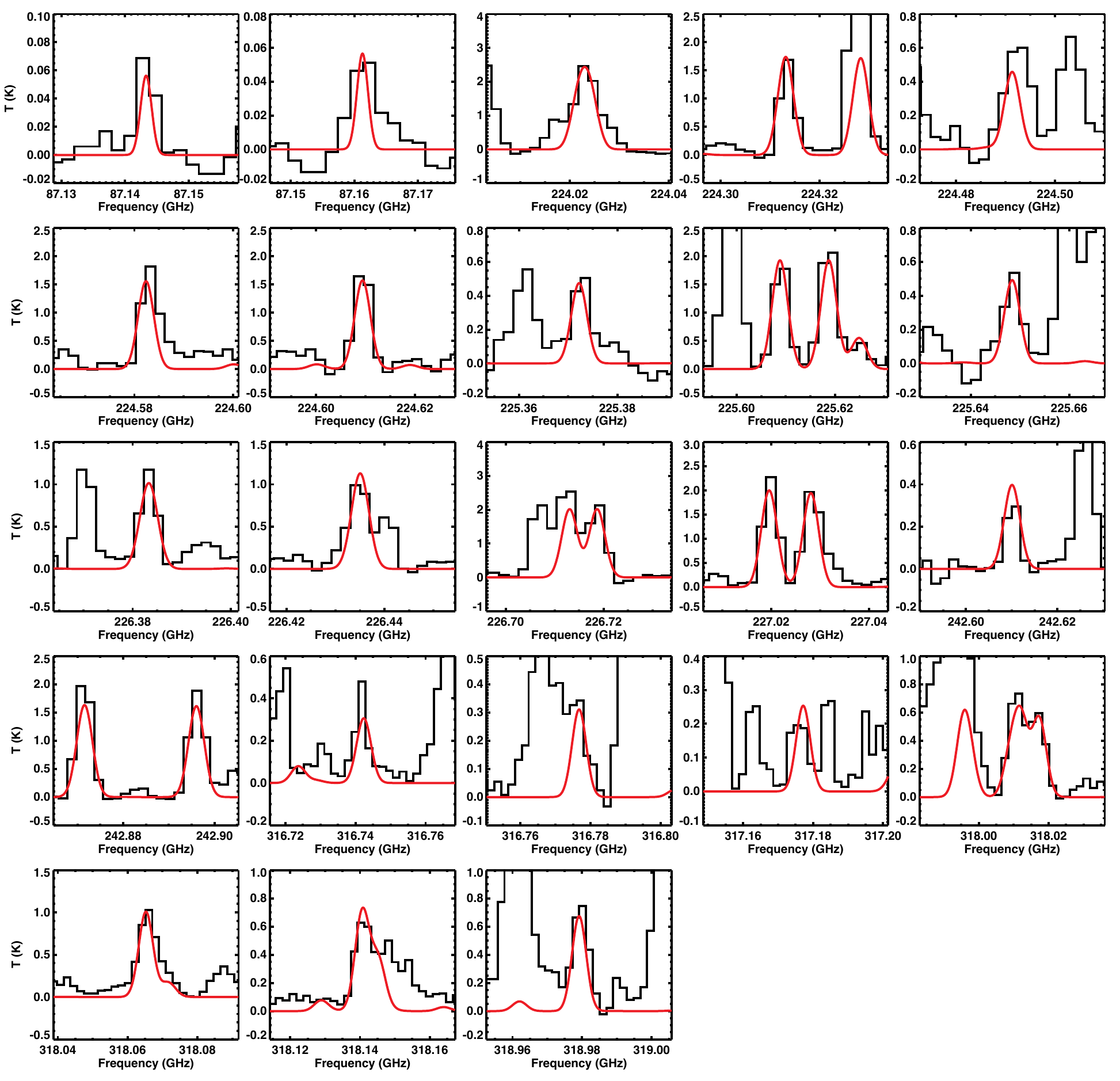}
\caption{
Observed lines of CH$_3$OCHO towards the protostar NGC~1333~IRAS2A (in black). The best-fit model for CH$_3$OCHO (see Table \ref{table_model})} is shown in red lines. 
\label{Model_CH3OCHO}
\end{center}
\end{figure*}

We carried out a local thermodynamic equilibrium analysis of the three species through the rotational diagram method \citep{Goldsmith1999}. We consider that the lines are emitted in a region of 0.5$\arcsec$ size, which is the average size derived for the methyl formate lines when fitting circular gaussians in the ($u$,$\varv$) plane (see also \citealt{Maury2014}). It is also similar to what we found for deuterated water \citep{Coutens2014}. It also corresponds to the expected size of the region where the temperature increases above $\sim$100\,K according to dust radiation transfer models of the envelope \citep{Jorgensen2002} and where the complex molecules and deuterated water should sublimate from the grains.
The line fluxes that result from a combination of several transitions of the same species are used in the rotational diagrams, unless the transitions have different $E_{\rm up}$ values. For glycolaldehyde, we include two lines slightly blended with some methyl formate transitions after subtraction of the predicted flux contribution from methyl formate. As the best-fit model for methyl formate reproduces extremely well the observations, the final fluxes of the glycolaldehyde lines can be trusted, which is also confirmed by their alignment with the other points in the rotational diagram of glycolaldehyde (see upper panel in Figure \ref{Fig_RD}).
Assuming a source size of 0.5$\arcsec$, we derive column densities (with 1$\sigma$ uncertainties) of 2.4$^{+0.6}_{-0.3}$\,$\times$\,10$^{15}$\,cm$^{-2}$, 1.3$^{+0.1}_{-0.1}$\,$\times$\,10$^{16}$\,cm$^{-2}$, and 4.8$^{+0.3}_{-0.3}$\,$\times$\,10$^{16}$\,cm$^{-2}$, and excitation temperatures of 103$^{+60}_{-20}$\,K, 133$^{+23}_{-14}$\,K, and 126$^{+9}_{-7}$\,K for glycolaldehyde, ethylene glycol, and methyl formate, respectively. Within the uncertainty range, the excitation temperature seems to be similar between the three species ($\sim$\,130\,K), which is consistent if the three species arise from a same region. 
We checked, for each species, that there is no line flux overpredicted by the model anywhere in the four datasets. 
In the case of ethylene glycol, the model shows an overproduced flux for some transitions, especially the lines (240.778, 241.545, 241.860, and 316.444 GHz) that correspond to the four lower points in the rotational diagram (see middle panel in Figure \ref{Fig_RD}). 
A model with a column density of 1.1\,$\times$\,10$^{16}$\,cm$^{-2}$ would be sufficient to produce line fluxes consistent with these observations.
Table \ref{table_model} summarizes the parameters used for the line modeling of the three species that can be seen in Figures \ref{Model_GA}, \ref{Model_aGgglycol}, and \ref{Model_CH3OCHO}.
According to these models, all lines are optically thin ($\tau$ $\leq$ 0.1).

\begin{table}[t!]
\caption{Parameters used to compute the synthetic spectra of glycolaldehyde, ethylene glycol, and methyl formate. }
\begin{center}
\label{table_model}
\begin{tabular}{lcccc}
\hline
\hline
Molecule & Source  & $T_{\rm ex}$ & $N$ & $\varv_{\rm LSR}$ \\
& size ($\arcsec$) & (K) & (cm$^{-2}$) & (km\,s$^{-1}$) \\
\hline
CH$_2$OHCHO & 0.5 & 130 & 2.4\,$\times$\,10$^{15}$ & 7.0 \\
aGg$'$-(CH$_2$OH)$_2$ & 0.5 & 130 & 1.1\,$\times$\,10$^{16}$ & 7.0 \\
CH$_3$OCHO & 0.5 & 130 & 4.8\,$\times$\,10$^{16}$ & 7.0 \\
\hline
\end{tabular}
\tablefoot{The FWHM used for the line modeling are 4.5, 5.0, 5.0, and 7.0 \,km\,s$^{-1}$ for the data at 317, 242, 225, and 86 GHz, respectively.}
\end{center}
\end{table}

Although no other species than glycolaldehyde is found at a frequency of 240890.5 MHz, the line ($E_{\rm up}$ = 52\,K) is probably blended with an unidentified species: the predicted flux is completely underproduced with respect to the observations, and it cannot be due to a different excitation in the cold gas, as a line of glycolaldehyde  at 243232.21 MHz ($E_{\rm up}$ = 47\,K) -- blended with a bright CH$_2$DOH line in the red-shifted part of the spectrum and also potentially blended with a DCOOH line ($E_{\rm up}$\,=\,106\,K, $A_{\rm ij}$\,=\,1.35\,$\times$\,10$^{-4}$\,s$^{-1}$) -- would have a higher flux inconsistent with the observed one.

\begin{table*}
\caption{(CH$_2$OH)$_2$/CH$_2$OHCHO, CH$_3$OCHO/CH$_2$OHCHO, and CH$_3$OCHO/(CH$_2$OH)$_2$ column density ratios determined in different objects. }
\begin{center}
\label{table_comp}
\begin{tabular}{lcccc}
\hline
\hline
Source & (CH$_2$OH)$_2$/CH$_2$OHCHO\tablefootmark{a} & CH$_3$OCHO/CH$_2$OHCHO & CH$_3$OCHO/(CH$_2$OH)$_2$\tablefootmark{a} & References \\
\hline
& \multicolumn{3}{c}{Class 0 protostars} \\
\cline{2-5}
NGC~1333 IRAS2A & $\sim$5 & $\sim$20 & $\sim$4 & 1 \\
IRAS~16293-2422 & $\sim$1 & $\sim$13  & $\sim$13 & 2 \\ 
\hline
& \multicolumn{3}{c}{Comets} \\
\cline{2-5}
C/1995 O1 (Hale-Bopp) & $\ge$ 6 & $\ge$ 2 &$\sim$ 0.3 & 3 \\
C/2012 F6 (Lemmon) &  $\ge$ 3 & ... & $\le$ 0.7 & 4 \\ 
C/2013 R1 (Lovejoy) & $\ge$ 5 & ... & $\le$ 0.6 & 4 \\
\hline
& \multicolumn{3}{c}{High-- and intermediate--mass star forming regions} \\
\cline{2-5}
Sgr B2(N)  & 0.7--2.2\tablefootmark{b} & $\sim$52\tablefootmark{c} & $\sim$30 & 5, 6, 7 \\
G34.41+0.31 & ... & $\le$ 34 & ... & 8 \\
NGC7129~FIRS2 & $\sim$2 & $\sim$40 & $\sim$20 & 9 \\
\hline
& \multicolumn{3}{c}{Molecular clouds in the Central Molecular Zone} \\
\cline{2-5}
G--0.02, G--0.11, and G+0.693 & $\sim$1.2--1.6 & $\sim$3.3--5.2 & $\sim$2.5--4.3 & 10 \\
\hline
\end{tabular}
\tablefoot{
\tablefoottext{a}{(CH$_2$OH)$_2$ refers to the aGg$'$ conformer.}
\tablefoottext{b}{Large scale emission.}
\tablefoottext{c}{Hot core emission. }}
\tablebib{1) this study, 2) J\o rgensen et al. (2012, in prep.), 3) \citet{Crovisier2004}, 4) \citet{Biver2014}, 5) \citet{Hollis2001}, 6) \citet{Hollis2002}, 7) \citet{Belloche2013}, 8) \citet{Beltran2009}, 9) \citet{Fuente2014}, 10) \citet{Requena2008}.}
\end{center}
\end{table*}

\section{Discussion}
\label{sect_discussion}

The relative abundances of the three species are derived from the column densities in Table \ref{table_model} and compared with other star-forming regions and comets in Table~\ref{table_comp}. 
The (CH$_2$OH)$_2$/CH$_2$OHCHO abundance ratio of $\sim$0.3--0.5 previously derived in IRAS16293 by \citet{Jorgensen2012} was revised. Indeed, the assignment in \citet{Jorgensen2012} was based on only one line of the gGg$'$ conformer of ethylene glycol about 200 cm$^{-1}$ ($\sim$290 K, \citealt{Muller2004}) above the lowest-energy aGg$'$ conformer -- and thus tentative. An analysis from observations of 6 transitions of the lower energy conformer from ALMA Cycle~1 observations at 3~mm (4 spectral windows at 89.48--89.73, 92.77--93.03, 102.48--102.73 and 103.18--103.42~GHz; J\o rgensen et al. in prep.) results in a higher ethylene glycol-to-glycolaldehyde abundance ratio of 1.0\,$\pm$\,0.3. 
This new estimate is consistent with the ratio expected between the aGg$'$ and gGg$'$ conformers under thermal equilibrium conditions at 300\,K, the excitation temperature of glycolaldehyde derived in IRAS16293 \citep{Jorgensen2012}.
The (CH$_2$OH)$_2$/CH$_2$OHCHO abundance ratio in IRAS2A is estimated at 5.5\,$\pm$\,1.0 if we consider the column densities derived from the rotational diagrams. It is however slightly lower (4.6) if we use the column density of ethylene glycol of 1.1\,$\times$\,10$^{16}$ cm$^{-2}$ that does not overproduce the peak intensities of a few lines (see Fig. \ref{Model_aGgglycol}). The (CH$_2$OH)$_2$/CH$_2$OHCHO abundance ratio is consequently a factor $\sim$5 higher than in the Class 0 protostar IRAS16293. It is also higher than in the other star-forming regions (see Table \ref{table_comp}), 
but comparable to the lower limits derived in comets ($\gtrsim$\,3--6). This indicates that the glycolaldehyde chemistry may vary among hot corinos in general. It is  possible that, like IRAS2A, other very young low-mass protostars show high (CH$_2$OH)$_2$/CH$_2$OHCHO abundance ratios, in agreement with the cometary values.
The CH$_3$OCHO/CH$_2$OHCHO column density ratio found in IRAS2A ($\sim$\,20) ranges between the values derived in the molecular clouds from the Galactic Center ($\sim$\,3.3--5.2) and the high-mass star-forming regions ($\sim$\,40--52). A lower limit of 2 was derived for the Hale-Bopp comet.

In contrast to IRAS16293, the (CH$_2$OH)$_2$/CH$_2$OHCHO abundance ratio in IRAS2A is comparable to the lower limits in comets.
To explain these different abundance ratios in IRAS2A and IRAS16293, two scenarios are possible: either the (CH$_2$OH)$_2$/CH$_2$OHCHO ratio is similar in the grain mantles of low-mass protostars and it evolves in the gas phase after the sublimation of the molecules in the hot corinos, or this ratio was already different in the grain mantles of the two protostars. 

In the first scenario, if we assume that the (CH$_2$OH)$_2$/CH$_2$OHCHO increases until it reaches the cometary value, it would mean that glycolaldehyde can easily be destroyed in the gas phase of the warm inner regions. Another possibility would be that ethylene glycol can form efficiently in the gas phase, but complex organic molecules are generally difficult to form with high abundances in the gas phase. If the evaporation temperature of ethylene glycol is higher than glycolaldehyde, as assumed in the chemical model of \citet{Garrod2013}, ethylene glycol would desorb later than glycolaldehyde and the (CH$_2$OH)$_2$/CH$_2$OHCHO abundance ratio would consequently increase with time (until the two molecules have completely desorbed).  
This chemical model however predicts an abundance of glycolaldehyde significantly higher than those of ethylene glycol and methyl formate, which is inconsistent with the ratios derived in IRAS2A. More theoretical and experimental work would be needed to make the case that these hypotheses are plausible.

\begin{figure}[t!]
\begin{center}
\includegraphics[scale=0.33]{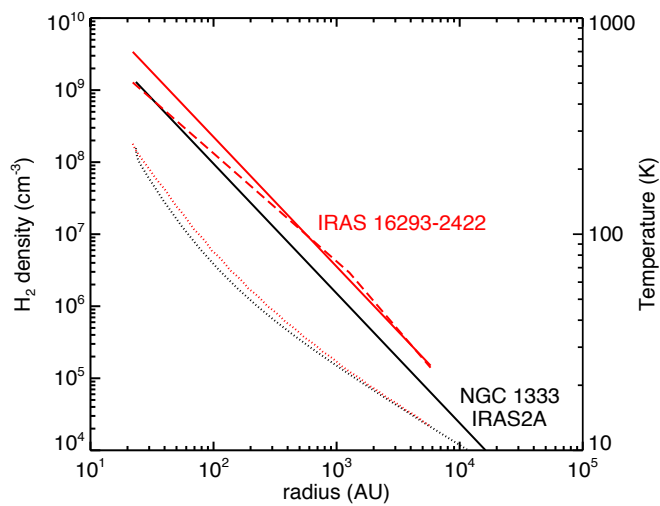}
\caption{Black: H$_2$ density (solid line) and temperature (dotted line) profiles of the protostar IRAS2A from \citet{Jorgensen2002}. Red: H$_2$ density (solid line: power-law model, dashed line: Shu-like model) and temperature (dotted line) profiles of the protostar IRAS16293 from \citet{Crimier2010}.} 
\label{Fig_comp_profile}
\end{center}
\end{figure}

In contrast, experimental studies based on irradiation of ices show that the second scenario \emph{is} likely.
Such studies show that glycolaldehyde, ethylene glycol, and methyl formate can be synthesized by irradiation of pure or mixed methanol (CH$_3$OH) ices \citep{Hudson2000,Oberg2009}. Interestingly, the (CH$_2$OH)$_2$/CH$_2$OHCHO abundance ratio is found to be dependent on the initial ice composition as well as the ice temperature during the UV irradiation. The CH$_3$OH:CO ratio in the ices is a key parameter: for irradiated 20~K ices a composition of pure CH$_3$OH leads to a (CH$_2$OH)$_2$/CH$_2$OHCHO ratio higher than 10, while a CH$_3$OH:CO 1:10 ice mixture produces a (CH$_2$OH)$_2$/CH$_2$OHCHO ratio lower than 0.25 \citep{Oberg2009}. The difference found between IRAS16293 and IRAS2A could then be related to a different grain mantle composition in the two sources. 
If the CH$_3$OH:CO ratio in the grain mantles of IRAS 2A was higher than in IRAS 16293, a higher (CH$_2$OH)$_2$/CH$_2$OHCHO abundance ratio would be expected according to the laboratory results. In fact, the CH$_3$OH gas-phase abundance in the inner envelope is found to be higher in IRAS2A ($\sim$\,4\,$\times$\,10$^{-7}$, \citealt{Jorgensen2005b}) than in IRAS16293 ($\sim$\,1\,$\times$\,10$^{-7}$, \citealt{Schoier2002}), while the CO abundance is relatively similar ($\sim$\,(2--3)\,$\times$\,10$^{-5}$, \citealt{Jorgensen2002,Schoier2002}). This could consequently be the result of the desorption of ices with a higher CH$_3$OH:CO ratio in IRAS2A than IRAS16293. The question then arises: how can CH$_3$OH be more efficiently produced on grains in IRAS2A than in IRAS16293? 
Several scenarios are possible:
\textit{i)} The initial conditions may play an important role in the CH$_3$OH:CO ratio. In particular, experiments and simulations show that the efficiency of CH$_3$OH formation through CO hydrogenation on the grains is dependent on temperature, ice composition (CO:H$_2$O), and time \citep{Watanabe2004,Fuchs2009}. 
\textit{ii)} The collapse timescale was longer in IRAS2A than in IRAS16293, enabling to form more CH$_3$OH. 
\textit{iii)} The H$_2$ density in the prestellar envelope of IRAS2A was lower than that of IRAS16293. Indeed a less dense environment would lead to a higher atomic H density and consequently to a higher efficiency of CO hydrogenation. This was proposed by \citet{Maret2004} and \citet{Bottinelli2007} to explain an anti-correlation found between the inner abundances of H$_2$CO and CH$_3$OH and the submillimeter luminosity to bolometric luminosity ($L_{\rm smm}$/$L_{\rm bol}$) ratios  of different low-mass protostars. The $L_{\rm smm}$/$L_{\rm bol}$ parameter is interpreted as an indication of different initial conditions, rather than an evolutionary parameter in this context \citep{Maret2004}. 
The $L_{\rm smm}$/$L_{\rm bol}$ ratios of IRAS2A ($\sim$\,0.005, \citealt{Karska2013}) and IRAS16293 ($\sim$\,0.019, \citealt{Froebrich2005}) are consistent with this hypothesis. The current H$_2$ density profiles of these two sources are also in agreement with this scenario, if they keep the memory of the prestellar conditions. The density derived in the outer envelope of IRAS2A with a power-law model \citep{Jorgensen2002} is lower than the density derived in IRAS16293 by \citet{Crimier2010} whether it be for a Shu-like model or a power-law model, while the temperature profiles are relatively similar (see Figure \ref{Fig_comp_profile}).
Along the same lines, \citet{Hudson2005} showed with proton irradiation experiments that glycolaldehyde is more sensitive to radiation damage than ethylene glycol. Irradiation would be more important in less dense envelopes, which would also be consistent with a less dense prestellar envelope in IRAS2A. 
A recent experiment by Fedoseev et al. (submitted) shows that these two species can also be synthesized by surface hydrogenations of CO molecules in dense molecular cloud conditions. They do not directly form from CH$_3$OH, but the results of this experiment show that similarly to CH$_3$OH that results from  successive hydrogenations of CO, ethylene glycol forms by two successive hydrogenations of glycolaldehyde. This is consequently in agreement with the proposed scenario.

In conclusion, the (CH$_2$OH)$_2$/CH$_2$OHCHO abundance ratio measured in low-mass protostars can be different from one source to another, and possibly consistent with cometary values. In some cases, the (CH$_2$OH)$_2$/CH$_2$OHCHO ratios determined in comets could consequently be inherited from early stages of star formation. Such a difference between low-mass protostars could be related to a different CH$_3$OH:CO ratio in the grain mantles. A more efficient hydrogenation (due for example to a lower density) on the grains would lead to higher abundances of CH$_3$OH and (CH$_2$OH)$_2$. A determination of (CH$_2$OH)$_2$/CH$_2$OHCHO ratios in larger samples of star-forming regions could help understand how the initial conditions (density, molecular cloud, ...) affect their relative abundances.

\begin{acknowledgements}
The authors are grateful to the IRAM staff, especially Tessel van der Laan, Arancha Castro-Carrizo, Chin-Shin Chang, and Sabine K{\"o}nig, for their help with the calibration of the data. This research was supported by a Junior Group Leader Fellowship from the Lundbeck Foundation (to JKJ). Centre for Star and Planet Formation is funded by the Danish National Research Foundation. MVP acknowledges EU FP7 grant 291141 CHEMPLAN. The research leading to these results has received funding from the European Commission Seventh Framework Programme (FP/2007-2013) under grant agreement N$^{\rm o}$ 283393 (RadioNet3). 
\end{acknowledgements}

\end{document}